\documentclass[useAMS,usenatbib]{mn2e}
\usepackage{graphicx}
\usepackage{color}
\usepackage{amsmath}

\title[A treatment procedure for GMOS/IFU data cubes]{A treatment procedure for GMOS/IFU data cubes: application to NGC 2835}

\author[Menezes et al.] {R.~B.~Menezes$^{1,2}$\thanks{E-mail:
roberto.menezes@ufabc.edu.br}, T.~V.~Ricci$^3$, J.~E.~Steiner$^1$, Patr\'icia~da~Silva$^1$, Fabricio~Ferrari$^{4}$
\newauthor and B.~W.~Borges$^5$ \\
$^{1}$Instituto de Astronomia Geof\'isica e Ci\^encias Atmosf\'ericas, Universidade de S\~ao Paulo, S\~ao Paulo, 05508-090, SP, Brazil \\
$^{2}$Centro de Ci\^encias Naturais e Humanas, Universidade Federal do ABC, Santo Andr\'e, 09210-580, SP, Brazil \\
$^{3}$Universidade Federal da Fronteira Sul, Cerro Largo, 97900000, RS, Brazil \\
$^{4}$Instituto de Matem\'atica Estat\'istica e F\'isica, Universidade Federal do Rio Grande, Rio Grande, 96201900, RS, Brazil \\
$^{5}$ Coordenadoria Especial de F\'isica, Qu\'imica e Matem\'atica, Universidade Federal de Santa Catarina, Ararangu\'a, 88905120, SC, Brazil}

\begin{document}

\date{Accepted 2018 November 21}

\pagerange{\pageref{firstpage}--\pageref{lastpage}} \pubyear{2018}

\maketitle

\label{firstpage}

\begin{abstract}
We present a set of treatment techniques for GMOS/IFU data cubes, including: correction of the differential atmospheric refraction; Butterworth spatial filtering, to remove high spatial-frequency noise; instrumental fingerprint removal; Richardson-Lucy deconvolution, to improve the spatial resolution of the observations. A comparison with HST images shows that the treatment techniques preserve the morphology of the spatial structures in the data cubes, without introducing distortions or ghost features. The treatment procedure applied to the data cube of the central region of the late-type galaxy NGC 2835 allowed the detection of a nuclear emitting region with emission-line ratios indicating a probable LINER, although the classifications of Transition Object or Seyfert galaxy cannot be discarded. Such a detection was not possible without the data treatment. This shows that the benefits provided by this treatment procedure are not merely cosmetic, but actually affect the quality of different types of analyses, like the determination of the fraction of galaxies, of different morphological types, with a central Seyfert/LINER/Transition object emission.

\end{abstract}


\section{Introduction}

Analyses of astronomical objects involving both spatial and spectral information have become possible with the introduction of data cubes, which are data sets with two spatial dimensions and one spectral dimension. Data cubes show images at different wavelengths or spectra of different spatial regions and can be obtained with instruments like Integral Field Spectrographs or Fabry-Perot spectrographs. Two examples of Integral Field Spectrographs are the Spectroscopic Areal Unit for Research on Optical Nebulae (SAURON; Bacon et al. 2001) and the Pmas fibre PAcK (PPAK) Integral Field Unit (IFU), used for the Calar Alto Legacy Integral Field Area (CALIFA; S\'anchez et al. 2012). Data cubes contain a large amount of information but can also contain instrumental effects like high spatial-frequency noise, for example. Such effects can have a significant impact on the results obtained with the performed analyses and, therefore, should be removed.

\begin{table*}
\begin{center}
\caption{Details of the observations used in the paper to present the treatment procedure.\label{tbl1}}
\begin{tabular}{cccccccc}
\hline
Target  &  Programme ID  &  Name of PI  &  Observation date  &  Grating  &  Exposure time (s)  &  Number       &   Seeing \\
            &                           &                      &                                &            &                                &  of exposures   &  (arcsec) \\
\hline
HZ 44  &  GN-2012A-Q-12  &  H. -Y. Shih  &  2012 April 24  &  B600+G5307  &  300.0  &  1      &     0.73\\
LTT 1020  &  GS-2012B-Q-52  &  R. B. Menezes  &  2012 November 28  &  B600+G5323  &  900.5  &  1 & 1.16 \\
LTT 1020  &  GS-2013B-Q-20  &  T. V. Ricci  &  2013 September 25 &  B600+G5323  &  900.5  &  1  &   0.75  \\
LTT 2415  &  GS-2014B-Q-30  &  J. E. Steiner  &  2014 November 28  &  R831+G5322  &  300.0  &  1  &  0.55  \\
LTT 3218  &  GS-2014A-Q-5  &  J. E. Steiner  &  2014 March 14  &  R831+G5322  &  300.5  &  1  &  1.37  \\
LTT 7379  &  GS-2013A-Q-82  &  R. B. Menezes  &  2013 July 31  &  R831+G5322  &  901.0  &  1  &  2.12  \\
LTT 9239  &  GS-2008B-Q-21  &  J. E. Steiner  &  2008 July 29  &  B600+G5323  &  180.5  &  1  &  1.12  \\
NGC 157  &  GS-2014B-Q-30  &  J. E. Steiner  &  2014 December 23  &  R831+G5322  &  930.0  &  3  &  0.49 \\
NGC 1313  &  GS-2012B-Q-52  &  R. B. Menezes  &  2012 December 4  &  B600+G5323  &  589.5  &  3  &  0.52  \\
NGC 1316  &  GS-2013B-Q-20  &  T. V. Ricci  &  2013 October 7  &  B600+G5323  &  1800.5  &  1  &  1.06  \\
NGC 1399  &  GS-2008B-Q-21  &  J. E. Steiner  &  2008 August 4  &  B600+G5323  &  1800.5  &  1  &  1.13  \\
NGC 2835  &  GS-2015A-Q-3    &  J. E. Steiner  &  2015 May 12    &  R831+G5322  &   865.0   &   3  &  0.40 \\
NGC 5236  &  GS-2014A-Q-5  &  J. E. Steiner  &  2014 February 23  &  R831+G5322  &  815.5  &  3  &  0.63  \\
NGC 7424  &  GS-2013A-Q-82  &  R. B. Menezes  &  2013 September 23  &  R831+G5322  &  808.5  &  3  &  0.54  \\
\hline
\end{tabular}
\end{center}
\end{table*}

The Gemini Multi-Object Spectrograph (GMOS) at the Gemini-North and Gemini-South telescopes provides long-slit and multi-slit spectral observations in the wavelength range of 3600 - 9400 \AA. Each of these instruments has also an IFU (Allington-Smith et al. 2002), which performs 3D spectroscopy, resulting in seeing-limited data cubes. The GMOS/IFU has two fields of view (FOVs): one for the science observation and the other for the sky observation. In the two-slit mode, the science FOV has 5 arcsec $\times$ 7 arcsec and is sampled by 1000 lenslets (each one with a diameter of 0.2 arcsec), while the sky FOV (at 1 arcmin from the science FOV) has 5 arcsec $\times$ 3.5 arcsec and is sampled by 500 lenslets. The incident light passes to 1000 fibres (each one connected to a lenslet) and is conducted to a group of slits, where it is dispersed and originates the spectra, which are recorded in the detector. The functioning of this instrument is entirely analogous in the one-slit mode but, in this case, the science FOV has 5 arcsec $\times$ 3.5 arcsec, being sampled by 500 lenslets, and the sky FOV has 5 arcsec $\times$ 1.75 arcsec, being sampled by 250 lenslets. The one-slit mode results in a spectral coverage with $\sim 2000 - 3000$ \AA, depending on the grating used for the observation and it has half of the spectral coverage of the one-slit mode.

This is the third and final paper of a series in which we describe a treatment methodology to be applied to data cubes. We presented treatment procedures for data cubes obtained with the Near-Infrared Integral Field Spectrograph (NIFS; McGregor et al. 2003) and with the Spectrograph for Integral Field Observations in the Near Infrared (SINFONI; Eisenhauer et al. 2003; Bonnet et al. 2004) in Menezes, Steiner \& Ricci (2014a, hereafter Paper I) and in Menezes et al. (2015, hereafter Paper II), respectively. In this work, we discuss the procedures used to treat data cubes obtained with the GMOS/IFU. Similarly to the procedures described in Paper I and Paper II, the methodologies presented in this work were applied using scripts written in Interactive Data Language\footnote{http://www.exelisvis.com/ProductsServices/IDL.aspx} ({\sc idl}). In Section 2, we give observational details of the data cubes used to illustrate the different treatment procedures in this work. In Section 3, we discuss the effect of the differential atmospheric refraction (DAR) on GMOS/IFU data cubes and how it can be removed. In Section 4, we present the Butterworth spatial filtering, used to remove high spatial-frequency noise from the data cubes. A characteristic `instrumental fingerprint', which is usually present in GMOS/IFU data cubes, together with the procedure used to remove it, is described in Section 5. In Section 6, we discuss the Richardson-Lucy deconvolution and the improvement in the spatial resolution of GMOS/IFU data cubes obtained with this procedure. In Section 7, we compare the spatial morphologies of the treated data cubes of the central regions of a few nearby galaxies with those of \textit{Hubble Space Telescope} (\textit{HST}) images of these objects. In Section 8, in order to provide a scientific example of the importance of our treatment procedures, we analyse the data cube of the nuclear region of the galaxy NGC 2835 before and after the treatment. In Section 9, we summarize the steps of the treatment and draw our conclusions.

\section{Observations and data reduction}

\begin{figure*}
\begin{center}
  \includegraphics[scale=2.0]{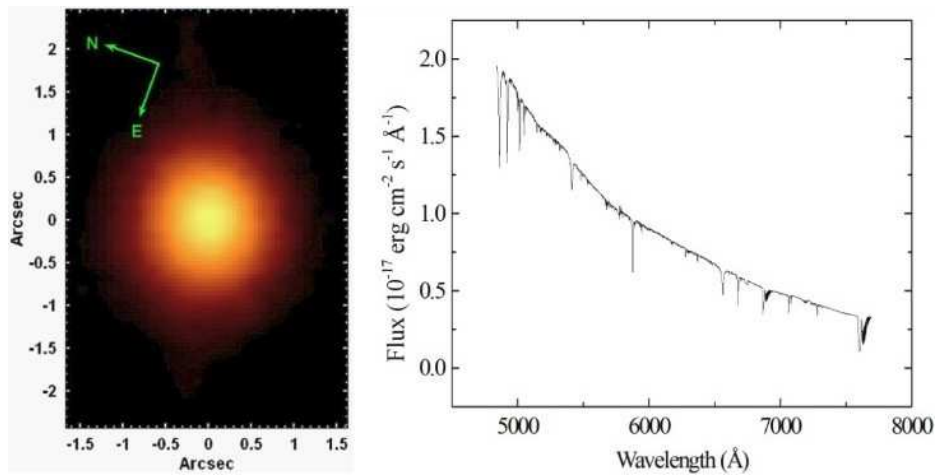}
  \caption{Average spectrum and image of the collapsed data cube of the star HZ 44.\label{fig1}}
\end{center}
\end{figure*}

The details of all the observations used in this paper are shown in Table~\ref{tbl1}. The data cube of the star HZ 44, retrieved from the Gemini Science Archive, is used throughout the paper to show the effects of the following treatment procedures: correction of the DAR, Butterworth spatial filtering and Richardson-Lucy deconvolution. The removal of the instrumental fingerprint, which is clearly visible only in data cubes of extended sources and not of point-like objects like stars, is illustrated using the data cubes of the central regions of the galaxies NGC 157, NGC 1316, and NGC 1399. In order to evaluate the reliability of our treatment methodology in preserving the spatial morphology of the objects, without introducing distortions in the images for example, we compared images of the data cubes of the central regions of NGC 1313, NGC 5236, and NGC 7424 with \textit{HST} images of these objects with the same FOV. Finally, as mentioned before, the analysis of the data cube of the central region of NGC 2835 was used as a scientific example demonstrating the importance of our treatment procedures. The standard star LTT 1020 was used for the flux calibration of the data cubes of NGC 1313 and NGC 1316, in two different programmes. The standard stars LTT 3218, LTT 7379, LTT 9239, and LTT 2415 were used for the flux calibration of the data cubes of NGC 5236, NGC 7424, NGC 1399, and NGC 157, respectively. 

Bias, GCAL-flat, twilight flat, and CuAr lamp images were obtained for the data reduction, which was performed in \textsc{IRAF} environment, with the \textsc{gemini} package. First, trimming and bias subtraction were performed. The cosmic rays were removed using the L. A. Cosmic routine \citep{van01}. After that, a correction of bad pixels was applied and the spectra were extracted. Then the spectra were corrected for gain variations along the spectral pixels, with response curves provided by the GCAL-flat images, and for gain variations along the fibres, with response maps obtained from the twilight flat images. The wavelength calibration was performed using the CuAr lamp images. In order to remove the sky emission, the average spectrum of the sky FOV was subtracted from the science data. Finally, using the observed standard stars, the data were flux calibrated, taking into account the atmospheric extinction, and the data cubes were constructed with spatial pixels (spaxels) of 0.05 arcsec. The collapsed image and the average spectrum of the reduced data cube of HZ 44 are shown in Fig.~\ref{fig1}.

\section{Correction of the DAR}

The correction of the DAR is the first step in the treatment of GMOS/IFU data cubes. As explained in Paper I, this effect generates a spatial displacement of the observed structures along the spectral axis of the data cubes. For a detailed description of this effect, see Paper I and also \citet{bon98} and \citet{fil82}. The DAR effect is considerably more intense in the optical than in the near infrared. As a consequence, the impact of this effect on GMOS/IFU data cubes is much more significant than on NIFS or SINFONI data cubes.

\begin{figure*}
\begin{center}
  \includegraphics[scale=0.3]{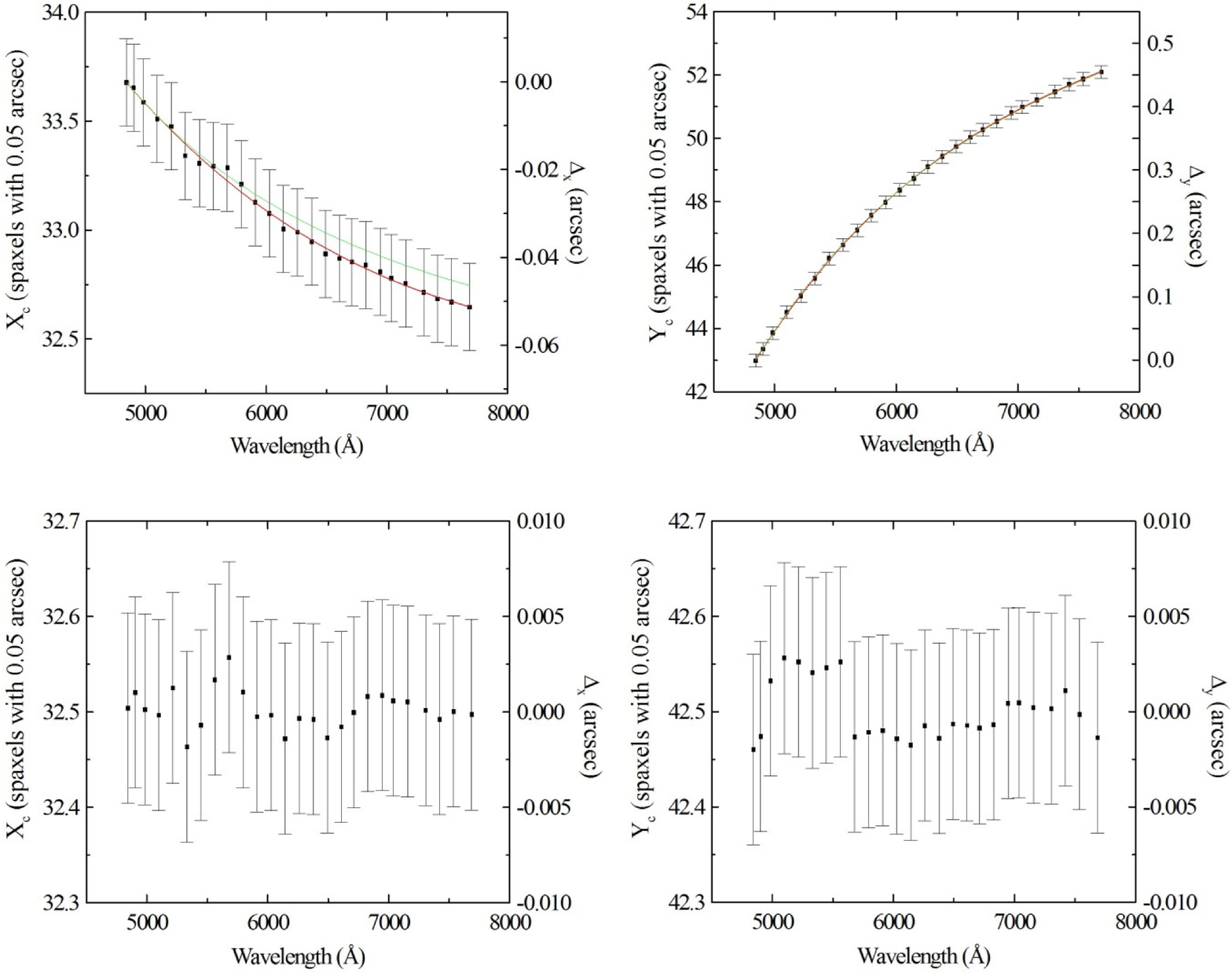}
  \caption{Graphs of the coordinates $X_c$ and $Y_c$ of the centre of the star HZ 44 for different wavelength values (top) before and (bottom) after the correction of the DAR effect. The red curves represent third-degree polynomials fitted to the points and the green curves correspond to the predictions obtained with the theoretical equations of the DAR effect, assuming a plane-parallel atmosphere.\label{fig2}}
\end{center}
\end{figure*}

\begin{figure}
\begin{center}
  \includegraphics[scale=1.25]{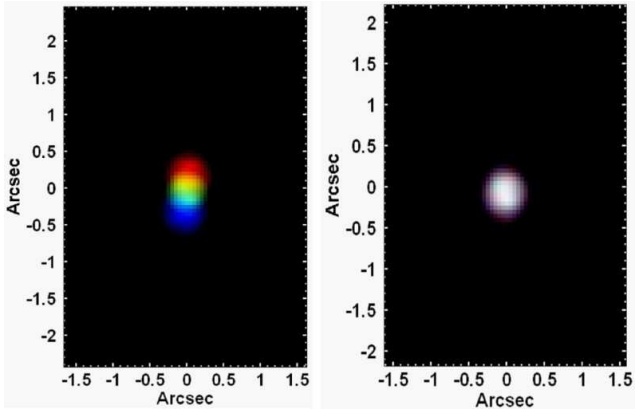}
  \caption{RGB composite images of the data cube of the star HZ 44 (left) before and (right) after the correction of the DAR effect. The blue, green and red colours correspond to images of the wavelength intervals $4840 - 4850$, $5905 - 5915$ and $7680 - 7690$ \AA, respectively.\label{fig3}}
\end{center}
\end{figure}

The star HZ 44 was observed at high airmass ($X = 1.303$). Fig.~\ref{fig2} shows graphs of the coordinates $X_c$ and $Y_c$ of the centre of the star HZ 44 as a function of the wavelength. These coordinates were obtained from images of the data cube at certain wavelength intervals, with 15\AA~~each. The wavelength values in the graphs in Fig.~\ref{fig2} correspond to the mean values in each one of these intervals. We can see that the total spatial displacement of the star HZ 44 is $\sim 0.05$ and $\sim 0.46$ arcsec along the horizontal and vertical axes, respectively. In nights with very good seeing conditions, point spread functions (PSFs) with full width at half-maximum (FWHM) values in the range of 0.4 - 0.5 arcsec can be obtained for GMOS/IFU data cubes. The spatial displacements caused by the DAR effect can be larger than the FWHM of the typical PSFs of such data cubes and, therefore, can affect the analysis. However, even in cases when the total spatial displacement caused by the DAR is smaller than the size of the PSF, it is advisable to apply the correction of this effect. The spatial displacement caused by the DAR in the data cube of HZ 44 can be seen very clearly in the RGB composite image in Fig.~\ref{fig3}, which shows images at different wavelength ranges.

\begin{figure}
\begin{center}
  \includegraphics[scale=1.9]{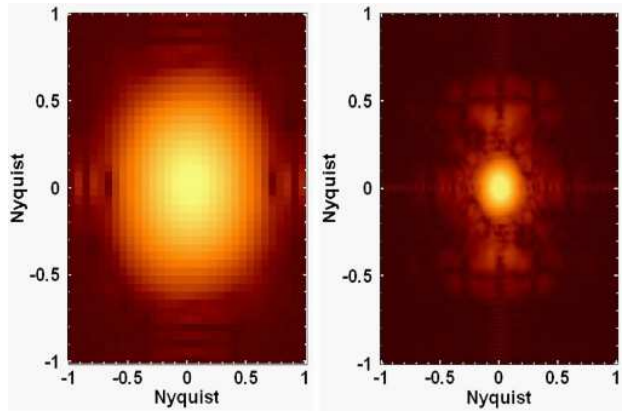}
  \caption{Moduli of the Fourier transforms of images, at an intermediate wavelength, of the data cubes of the star HZ 44 (left) with spaxels of 0.2 arcsec (without any spatial resampling) and (right) with spaxels of 0.05 arcsec (after the spatial resampling). High spatial-frequency components introduced by the resampling appear as bright areas farther from the centre of the image on the right.\label{fig4}}
\end{center}
\end{figure}

\begin{figure*}
\begin{center}
  \includegraphics[scale=3.0]{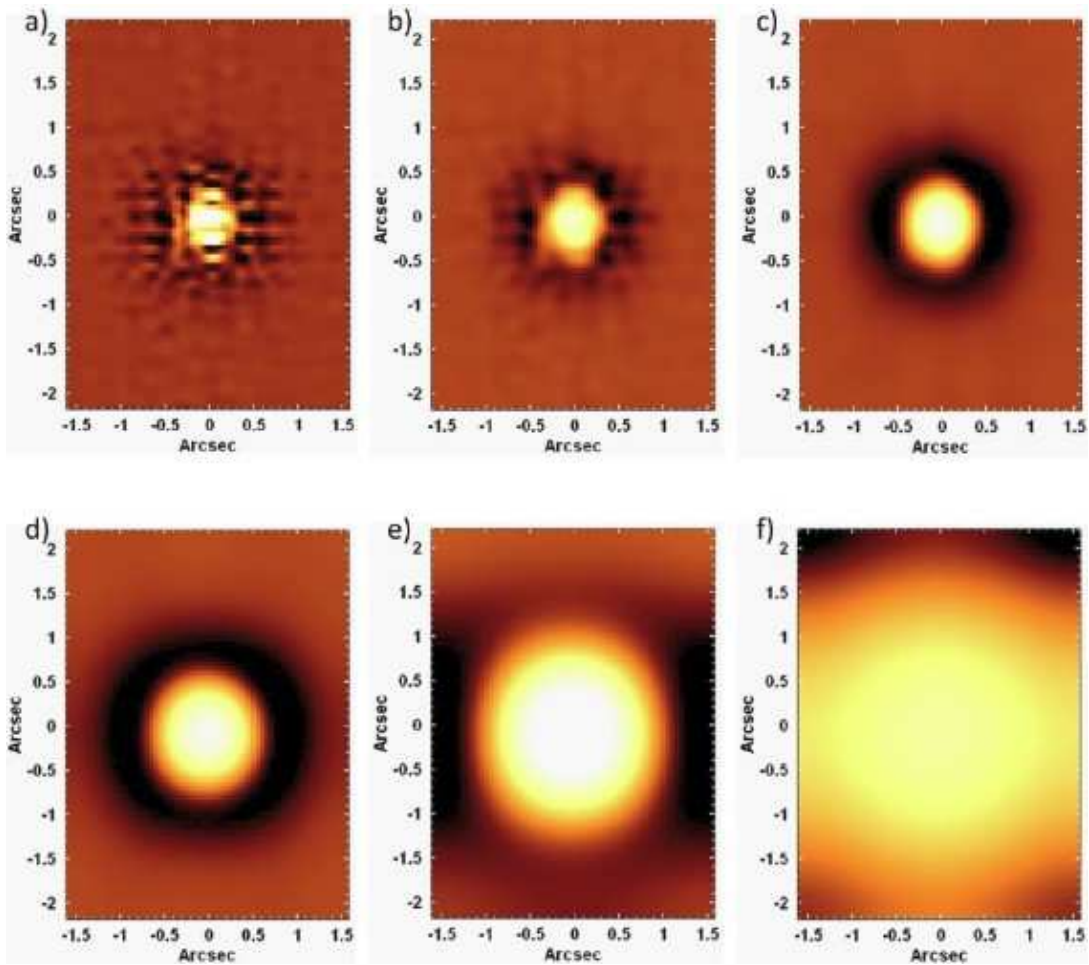}
  \caption{Images, at an intermediate wavelength, of the (a) $W_0$, (b) $W_1$, (c) $W_2$, (d) $W_3$, (e) $W_4$ and (f) $W_C$ data cubes of the star HZ 44. The high spatial-frequency components dominate the FOV of the $W_0$ data cube, but are essentially absent in the $W_2$ data cube.\label{fig5}}
\end{center}
\end{figure*}

\begin{figure*}
\begin{center}
  \includegraphics[scale=3.0]{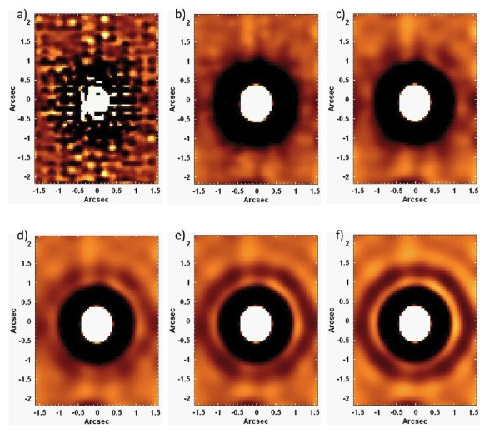}
  \caption{Images, at an intermediate wavelength, of the $W_0$ data cube of the star HZ 44 (a) before the Butterworth spatial filtering (the same as in Fig.~\ref{fig5} - a, but with a different look up table - LUT) and after the filtering with a cut-off frequency of 0.16 Ny and with (b) $n = 2$, (c) $n = 3$, (d) $n = 4$, (e) $n = 5$, and (f) $n = 6$. Structures similar to concentric rings appear in the data cubes filtered with $n \ge 2$.\label{fig6}}
\end{center}
\end{figure*}

As explained in Paper I and Paper II, we use two methods for removing the DAR effect: the practical and theoretical approaches. In the practical approach, the coordinates of the centres of a given spatial structure are measured as a function of the wavelength (as shown in the graphs in Fig.~\ref{fig2}). Then, third-degree polynomials are fitted to the points. Finally, we use a script that shifts each image of the data cube according to the values given by the fitted polynomials. In the theoretical approach, we use the equations of the DAR effect to apply the shifts to the images of the data cube. Such equations may assume a plane-parallel atmosphere \citep{bon98} or take into account the curvature of the Earth's atmosphere (task `\textit{refro}' from the {\sc `slalib'} package); however, we verified that, in the case of GMOS/IFU data cubes, the results obtained considering the curvature of the atmosphere are entirely compatible with those obtained with a plane-parallel atmosphere. So, for simplicity, we will only use the DAR theoretical equations in this work assuming a plane-parallel atmosphere. The third-degree polynomials fitted to the points corresponding to the coordinates of the centres of HZ 44, together with the predictions obtained with the theoretical equations, are shown in Fig.~\ref{fig2}. In the graph of $Y_c \times \lambda$, the theoretical curve and the fitted third-degree polynomial are nearly coincident and both reproduce precisely the behaviour of the observed points. In the graph of $X_c \times \lambda$, the theoretical curve is not exactly coincident with the fitted polynomial but it is compatible with the observed points, at a 1$\sigma$ level. We analysed the data cubes of many other stars observed with the GMOS/IFU and obtained similar results. Therefore, we conclude that both the theoretical and practical approaches are effective in reproducing the spatial displacements caused by the DAR in GMOS/IFU data cubes and can be used for the removal of this effect. It is worth mentioning that the practical approach can only be applied to data cubes in which the coordinates of a centroid can be determined with good precision.

We corrected the data cube of HZ 44 for the DAR effect, using the practical approach. The graphs of the coordinates $X_c$ and $Y_c$ as a function of the wavelength, after the correction of the DAR, are shown in Fig.~\ref{fig2}. The expected coordinates $X_c$ and $Y_c$ for the centre of HZ 44, after the correction, are 32.5 and 42.5, respectively; therefore, based on the graphs, it is easy to see the great efficiency of our method for correcting the DAR, which resulted in a precision, 1$\sigma$, of $\sim 1$ mas for $X_c$ and $\sim 2$ mas for $Y_c$. Such an efficiency can also be seen in the RGB composite image in Fig.~\ref{fig3}, with images at different wavelength ranges from the DAR corrected data cube. We verified that our algorithm for the DAR correction usually results in precisions between 1 and 10 mas for GMOS/IFU data cubes. It is important to mention that our purpose here is only to emphasize the importance to apply the correction of the DAR in GMOS/IFU data cubes before any analysis is performed. Although we used our algorithm to describe the procedure, there are other precise algorithms that can also be used. One example is the \textsc{gfcube} task in the \textsc{gemini} \textsc{iraf} package, which constructs the data cube, at the end of the reduction, and may also correct the effect of the DAR.

\section{Butterworth spatial filtering}

In situations when there is more than one exposure of the object that will be analysed, it is advisable to combine such exposures, in the form of a median, after the correction of the DAR effect, in order to remove possible cosmic rays or bad pixels that were not removed during the reduction. Since we retrieved only one exposure of the star HZ 44, this procedure was not applied. The next step in our treatment procedure of GMOS/IFU data cubes is the Butterworth spatial filtering \citep{gon02}, which removes, directly in the frequency domain, high spatial-frequency components from the images of the data cubes. As shown in equation (7) in Paper I, the Butterworth filter has two parameters: the cut-off frequency, $D_0$, and the order of the filtering, $n$. $D_0$ corresponds to the frequency at which the filter has a value of 0.5. On the other hand, the order $n$ indicates how steep is the profile of filter (the higher is $n$, the steeper is the filter). 

It would be intuitive to construct GMOS/IFU data cubes with spaxels of 0.2 arcsec, which is the same size of the lenslets that sample the science FOV of the instrument. However, as mentioned in Section 2, we constructed the data cubes with spaxels of 0.05 arcsec. In other words, a spatial resampling was applied during the construction of the data cubes, at the end of the reduction. The reason for this procedure is that smaller spaxels improve the visualization of the contours of the spatial structures. In addition, the Richardson-Lucy deconvolution results in better spatial resolutions when applied to images with smaller spaxels (see Section 6). As explained in Papers I and II, any spatial resampling introduces high-frequency components. Therefore, although most of the high spatial-frequency components observed in GMOS/IFU data cubes probably represent an instrumental effect, at least part of these structures may have been introduced by the spatial resampling. In order to evaluate that, we calculated the Fourier transforms of images, at an intermediate wavelength, of the data cubes of HZ 44 with spaxels of 0.2 arcsec (without any spatial resampling) and with spaxels of 0.05 arcsec (after the spatial resampling). The moduli of these Fourier transforms are shown in Fig.~\ref{fig4}. The high spatial-frequency components introduced by the resampling appear as bright areas farther from the centre of the image.

In order to visualize in more detail the high spatial-frequency components in the data cube of HZ 44, we applied a wavelet decomposition \citep{sta06}, using the \`A Trous algorithm. This process decomposes a signal in components with different frequencies. More specifically, the \`A Trous algorithm generates five components, which are usually called $w0$, $w1$, $w2$, $w3$, and $w4$ ($w0$ and $w4$ being the components with the highest and the lowest spatial frequencies, respectively), and also a smooth version of the original signal ($w_c$). The average of each component is 0 and the original signal can be recovered by adding all the wavelet components and $w_c$. In this case, the wavelet decomposition was applied to each image of the data cube, which resulted in six wavelet data cubes ($W_0$, $W_1$, $W_2$, $W_3$, $W_4$, and $W_C$). For more details of the wavelet decomposition, see Paper I. Fig.~\ref{fig5} shows images, at an intermediate wavelength, of the wavelet data cubes of HZ 44. The high spatial-frequency components are evident in the $W_0$ data cube and compromise significantly the visualization of the star. Such components are much less intense in the $W_1$ data cube and are essentially absent in the $W_2$ data cube. The dark ring, containing negative values, in the images is a typical feature of wavelet decompositions obtained with this method and tends to appear around every point-like source in the image. It is a natural consequence of the fact that the average of each wavelet component is 0.

\begin{figure*}
\begin{center}
  \includegraphics[scale=2.5]{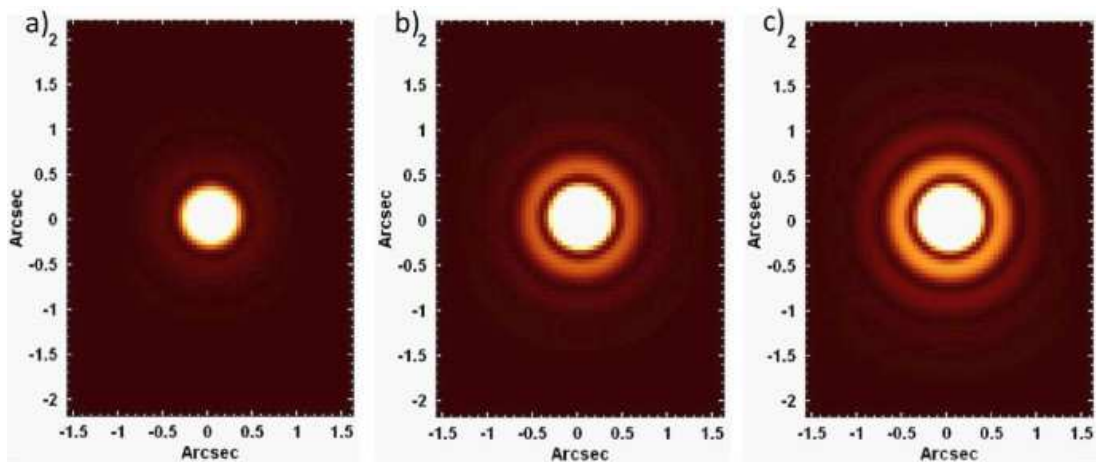}
  \caption{Moduli of the inverse Fourier transforms of circular Butterworth filters with a cut-off frequency of 0.16 Ny and (a) $n = 2$, (b) $n = 4$, and (c) $n = 6$. Concentric rings are significant in the inverse Fourier transforms of the filters with $n > 2$.\label{fig7}}
\end{center}
\end{figure*}

Considering the morphology of the Fourier transforms of the images of GMOS/IFU data cubes, we conclude that one possible approach is to apply the Butterworth spatial filtering using an elliptical filter, which is the same filter shape we adopted for the filtering of SINFONI data cubes (see Paper II). The general formula of an elliptical Butterworth filter, with cut-off frequencies of $a$ and $b$, along the horizontal and vertical axes, respectively, is given by equation (1) in Paper II. On the other hand, we also verified that the filtering process can be a lot simplified by assuming a simple circular filter ($a = b$), as the results do not show measurable differences relative to the ones obtained with $a \neq b$. 

In order to evaluate the most adequate value for the order of the filtering of GMOS/IFU data cubes, we applied the Butterworth spatial filtering to the $W_0$ data cube of HZ 44, using circular filters with a constant cut-off frequency of 0.16 Ny (as determined from the obtained Fourier transforms - Fig.~\ref{fig4}) and different values of $n$. It is worth mentioning that Ny is the Nyquist frequency, which is equal to half of the sampling frequency of the image. Images, at an intermediate wavelength, of the obtained filtered $W_0$ data cubes are shown in Fig.~\ref{fig6}. All the images reveal a considerable improvement of the visualization of the central star, due to high-frequency noise removed by the filtering procedure. On the other hand, higher values of $n$ generate concentric ``ring-like'' structures around the central star. These ``rings'' are almost absent for $n = 2$ but become more intense for $n \geq 3$. Similar features, but with a relatively different morphology, were also detected in Papers I and II, when the Butterworth spatial filtering, with high orders, was applied to NIFS and SINFONI data cubes, respectively.  The explanation for this behaviour is the same one that was discussed in these previous works and involves the convolution theorem, which states that the product between two functions in the frequency domain is equivalent to the convolution between these functions in the spatial domain. Therefore, the process of the Butterworth spatial filtering of a data cube is equivalent to the convolution between each image of the data cube and the inverse Fourier transform of the filter. Fig.~\ref{fig7} shows images corresponding to the moduli of the inverse Fourier transforms of the filters, with $n=2,4$ and $6$, used for the Butterworth spatial filtering of the $W_0$ data cube of HZ 44. The concentric rings appear very clearly in the inverse Fourier transforms of filters with $n > 2$. Therefore, we conclude that the concentric rings in the images of the $W_0$ data cubes filtered with orders higher than 2 are due to the fact that the inverse Fourier transforms of these high-order filters also show concentric rings. Since these inverse Fourier transforms, according to the convolution theorem, are convolved with the images of the data cube when the Butterworth spatial filtering is applied, the concentric rings end up being introduced in the images.

\begin{figure*}
\begin{center}
  \includegraphics[scale=2.3]{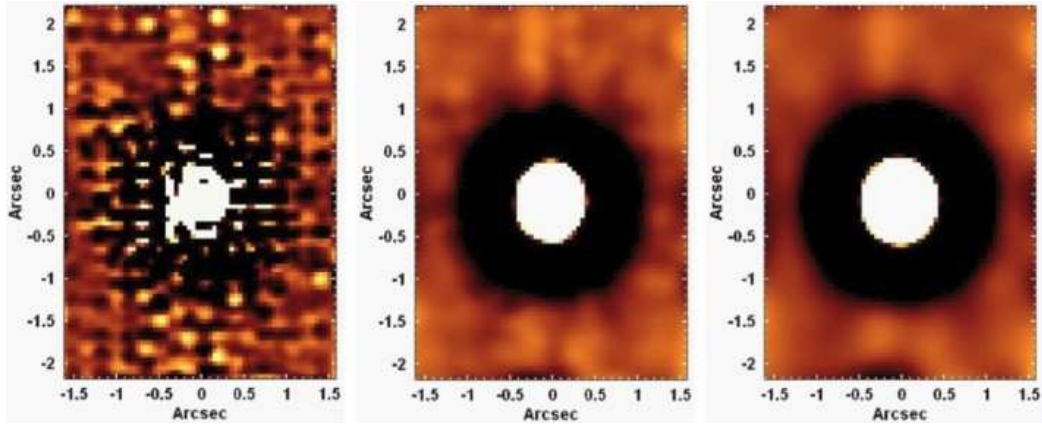}
  \caption{Images, at an intermediate wavelength, of the $W_0$ data cube of the star HZ 44 (right) before the Butterworth spatial filtering (the same as in Fig.~\ref{fig6} - a), (centre) after the filtering with a circular filter, with $n = 2$ and a cut-off frequency of 0.16 Ny, and (left) after the filtering with a squared circular filter, with $n = 2$ and a cut-off frequency of 0.18 Ny.\label{fig8}}
\end{center}
\end{figure*}

Based on the results above, the Butterworth spatial filtering with $n = 2$ is the most adequate to be applied to GMOS/IFU data cubes. However, one disadvantage of a low-order filter is that, due to the fact that its profile is less steep, it removes less noise than a high-order filter. As explained in Paper II, a solution for this problem is to use a filter given by the product of two identical filters with $n = 2$, which will enhance the noise removal, without introducing concentric rings. In this case, we opted to use the product of two identical circular filters (equation 2 in Paper II). Fig.~\ref{fig8} shows images, at an intermediate wavelength, of the $W_0$ data cube of HZ 44 filtered with a simple circular filter, with $n = 2$ and a cut-off frequency of 0.16 Ny, and with a squared circular filter, with $n =2$ and a cut-off frequency of 0.18 Ny. We can see that the filtering performed with a squared circular filter removed more noise, without introducing concentric rings. Therefore, we conclude that, similarly to the case of SINFONI data cubes, this approach is the most adequate to be used in GMOS/IFU data cubes. It should be noted that, as can be seen in Fig.~\ref{fig8}, the difference between the data cubes filtered with a simple circular filter and with a squared circular filter is subtle, so, although it is not an optimized approach, a simple circular filter, with $n = 2$, can also be used in the Butterworth spatial filtering of GMOS/IFU data cubes. The appropriate cut-off frequency of the filtering procedure changes considerably, depending on the PSF of the observation. Usually, values between 0.16 and 0.25 Ny produce good results, without compromising the PSF, in GMOS/IFU data cubes.

\begin{figure*}
\begin{center}
  \includegraphics[scale=1.9]{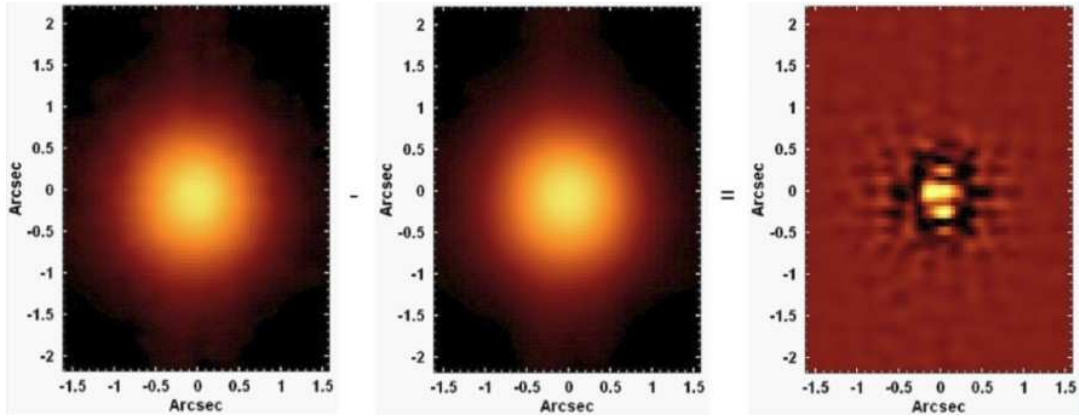}
  \caption{Left: image, at an intermediate wavelength, of the data cube of the star HZ 44 before the Butterworth spatial filtering. Centre: image, at an intermediate wavelength, of the data cube of the star HZ 44 after the Butterworth spatial filtering with a squared circular filter, with a cut-off frequency of 0.18 Ny and $n = 2$. Right: difference between the two previous images, showing the removed high spatial-frequency components.\label{fig9}}
\end{center}
\end{figure*}

Fig.~\ref{fig9} shows images, at an intermediate wavelength, of the original data cube of HZ 44 (without any wavelet decomposition) before and after the Butterworth spatial filtering. In order to show in detail the high spatial-frequency noise removed by the Butterworth spatial filtering, the difference between the images of the filtered and non-filtered data cubes is also shown in Fig.~\ref{fig9}. The morphology of the high spatial-frequency noise is considerably different from that of the noise observed in NIFS and SINFONI data cubes (Figs. 18 and 12 in Papers I and II, respectively), which appeared as vertical and horizontal stripes across the FOV. Since NIFS and SINFONI work with image slicers, which sample the FOV with slices along the vertical and horizontal axes, we believe that the high spatial-frequency noise detected in the data cubes obtained with these instruments is probably related to this mechanism. The GMOS/IFU FOV, on the other hand, is sampled by lenslets, each one connected to a fibre. Considering that and the morphology of the high spatial-frequency noise in Fig,~\ref{fig9}, we conclude that such noise is probably related to the lenslet/fibre mechanism of GMOS/IFU. Although a detailed evaluation about the origin of the high spatial-frequency noise in GMOS/IFU data cubes is beyond the scope of this paper, a few details about this topic should be discussed. The lenslets/fibres are packed in a way that there is no empty space between them. On the other hand, during the construction of the data cubes (at the end of the data reduction), in order to obtain flux values distributed along the spaxels in the images of the data cubes, it is necessary to interpolate the values, originally distributed according to the packing of the lenslets/fibres. Such interpolation can introduce high-frequency components and may be the cause of the observed high spatial-frequency noise.

\section{`Instrumental fingerprint' removal}

The GMOS/IFU data cubes usually show an instrumental feature that appears as vertical stripes across the FOV, with a low-frequency spectral signature. This instrumental fingerprint is considerably more intense than the one we detected in NIFS data cubes (Paper I) and is at least as intense as the fingerprint detected in SINFONI data cubes (Paper II), but with different spatial and spectral characteristics. Such characteristics are essentially the same for the fingerprints observed in the data cubes obtained with GMOS-North and GMOS-South. In the following sections, we explain the two methods we use for removing the instrumental fingerprint from GMOS/IFU data cubes. The first one is based on the Principal Component Analysis (PCA) Tomography technique \citep{ste09}, which consists of applying the PCA method \citep{mur87,fuk90} to data cubes. The second involves the wavelet decomposition, together with PCA Tomography (W-PCA Tomography). It is worth mentioning that the combination of the wavelet decomposition and PCA techniques has already been used in a previous study to remove high spatial-frequency noise from data cubes \citep{rif11}. However, the procedure we adopt here has certain differences and is focused on the removal of the instrumental fingerprint (and not the high spatial-frequency noise, which has already been removed with the Butterworth spatial filtering).

\subsection{Removal with PCA Tomography}

\begin{figure}
\begin{center}
  \includegraphics[scale=2.0]{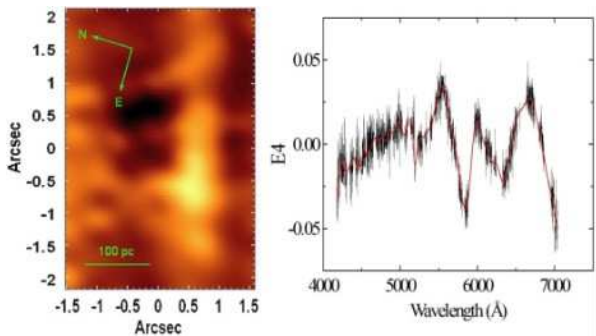}
  \caption{Eigenspectrum E4, showing the low-frequency pattern of the GMOS/IFU instrumental fingerprint (before the CCD replacement), and the corresponding tomogram, showing the spatial morphology of the fingerprint, obtained by applying PCA Tomography to the data cube of the central region of the galaxy NGC 1316, after the removal of the spectral lines. The spline fitted to the eigenspectrum is shown in red. The eigenspectrum and the tomogram show the spectral signature and the spatial morphology of the instrumental fingerprint, respectively.\label{fig10}}
\end{center}
\end{figure}

\begin{figure*}
\begin{center}
  \includegraphics[scale=2.5]{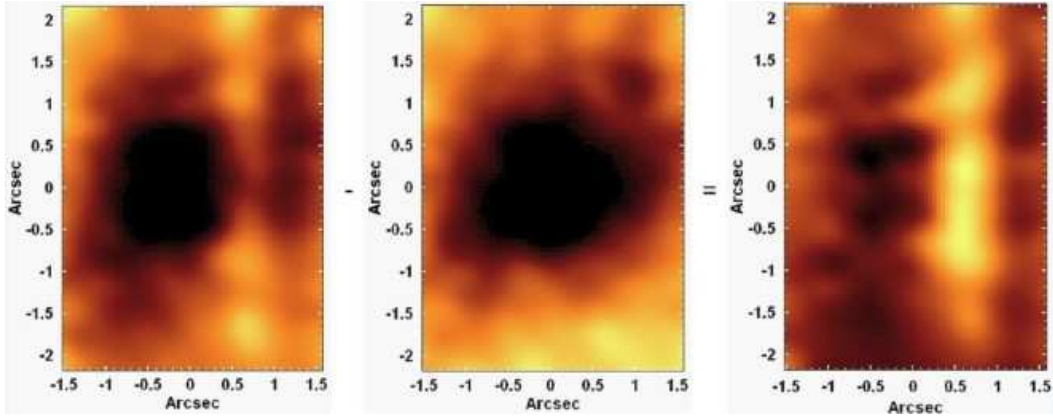}
  \caption{Left: image obtained, from the data cube of the central region of the galaxy NGC 1316 before the instrumental fingerprint removal, by subtracting the image of the wavelength interval $5652 - 5712$ \AA~from the image of the wavelength interval $5512 - 5572$ \AA. Centre: the same as in the image on the left, but from the data cube of NGC 1316 after the instrumental fingerprint removal. Right: difference between the two previous images. The structure corresponding to the instrumental fingerprint before the CCD replacement appears very clearly in the image on the left, but it is absent in the image at the centre.\label{fig11}}
\end{center}
\end{figure*}

This is the same approach described in Papers I and II to remove the instrumental fingerprint from NIFS and SINFONI data cubes, although the characteristics of the effect are somewhat different in this case. The first step of the procedure consists of removing the main absorption and emission lines from the data cube, which is necessary, as the fingerprint is related to the spectral continuum and not to the spectral lines. After that, the PCA Tomography technique is applied to the data cube without the spectral lines and the eigenvectors related to the fingerprint are identified. We then fit splines to the eigenspectra related to the fingerprint, taking into account only the low spectral-frequency pattern. Using these fitted splines and also the corresponding tomograms, we construct a data cube containing only the instrumental fingerprint. This last step is performed with the fitted splines, instead of the original eigenspectra, because we know that the fingerprint has a low-frequency spectral signature. Therefore, by constructing the data cube with these splines, we avoid that the result contains information associated with high spectral-frequency structures, which are not related to the instrumental fingerprint. Finally, the data cube containing only the fingerprint is subtracted from the original data cube.

\begin{figure}
\begin{center}
  \includegraphics[scale=1.8]{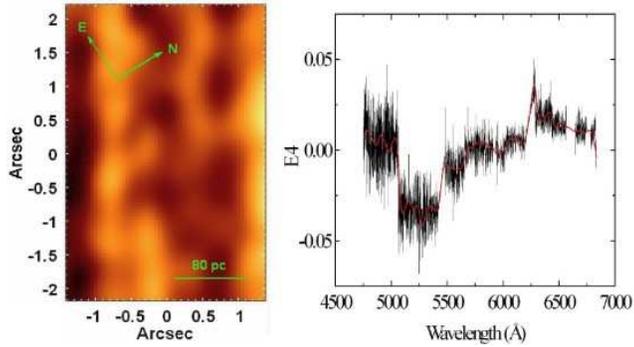}
  \caption{Eigenspectrum E4, showing the low-frequency pattern of the GMOS/IFU instrumental fingerprint (after the CCD replacement), and the corresponding tomogram, showing the spatial morphology of the fingerprint, obtained by applying PCA Tomography to the data cube of the central region of the galaxy NGC 157, after the removal of the spectral lines. The spline fitted to the eigenspectrum is shown in red. The spectral signature has a larger number of bumps than that in Fig.~\ref{fig10}. The number of vertical stripes in the tomogram is also higher than in Fig.~\ref{fig10}.\label{fig12}}
\end{center}
\end{figure}

\begin{figure*}
\begin{center}
  \includegraphics[scale=2.1]{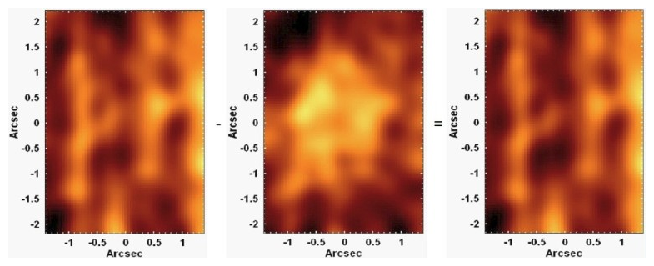}
  \caption{Left: image obtained, from the data cube of the central region of the galaxy NGC 157 before the instrumental fingerprint removal, by subtracting the image of the wavelength interval $5304 - 5401$ \AA from the image of the wavelength interval $6383 - 6480$ \AA. Centre: the same as in the image on the left, but from the data cube of NGC 157 after the instrumental fingerprint removal. Right: difference between the two previous images. The morphology of the spatial structures in the image on the left is clearly affected by the vertical stripes associated with the fingerprint. The removal of such stripes improved dramatically the visualization of the morphology of the spatial structures.\label{fig13}}
\end{center}
\end{figure*}

The GMOS-South and GMOS-North CCDs were replaced on 2014 June and 2017 February, respectively. We verified that such replacements changed the main aspects of the instrumental fingerprint of these spectrographs. In order to give an example of the instrumental fingerprint before the CCD replacement, we used a data cube of the central region of the galaxy NGC 1316, which was observed with GMOS-South. Fig.~\ref{fig10} shows eigenspectrum E4, together with the spline fitted to the eigenspectrum and the corresponding tomogram, obtained by applying PCA Tomography to the data cube of NGC 1316, after the removal of the main spectral lines. The low-frequency pattern of the fingerprint can be seen in the eigenspectrum, while the spatial morphology of the fingerprint, in the form of a broad vertical stripe, appears clearly in the tomogram. Different types of analyses can be affected by this instrumental fingerprint. One example of such an analysis involves the subtraction of images of different wavelength intervals in the data cube to identify regions with redder colours than others, which may be due to dust extinction. Fig.~\ref{fig11} shows the result obtained, from the data cube of NGC 1316 before the instrumental fingerprint removal, by subtracting the image of the wavelength interval $5652 - 5712$ \AA~from the image of $5512 - 5572$ \AA. Redder regions should appear dark in this image; however, the impact of the instrumental fingerprint in this analysis is very clear. The same result obtained from the data cube of NGC 1316 after the instrumental fingerprint removal is also shown in Fig.~\ref{fig11}. The improvement is evident, as the stripe corresponding to the fingerprint is not visible anymore. In order to provide more details about the spatial morphology of the fingerprint, Fig.~\ref{fig11} shows the difference between the images from the data cubes before and after the instrumental fingerprint removal. Another example of an analysis that can be significantly affected by this instrumental feature involves the spectral synthesis, which is often used to perform the stellar archaeology of an object.

A good example of the instrumental fingerprint after the CCD replacement is given by the data cube of the central region of the galaxy NGC 157. Fig.~\ref{fig12} shows eigenspectrum E4, together with the spline fitted to the eigenspectrum and the corresponding tomogram, provided by the PCA Tomography of the data cube of NGC 157, after the removal of the main spectral lines. We can see certain differences relative to the fingerprint before the CCD replacement. First, the spectral signature of the fingerprint in Fig.~\ref{fig12} has a larger number of bumps than that in Fig.~\ref{fig10}. The eigenspectrum in Fig.~\ref{fig12} also shows features similar to ``steps'' that are not clearly visible in the eigenspectrum in Fig.~\ref{fig10}. Besides that, the number of vertical stripes across the FOV is higher in the tomogram in Fig.~\ref{fig12} than in the tomogram in Fig.~\ref{fig10}. Following the approach adopted for the data cube of NGC 1316, Fig.~\ref{fig13} shows the results obtained, from the data cubes of NGC 157 before and after the instrumental fingerprint removal, by subtracting the image of the wavelength interval $5304 - 5401$ \AA~from the image of $6383 - 6480$ \AA. Again, the improvement provided by our procedure for removing the instrumental fingerprint is significant. The examples above involve data cubes obtained with GMOS-South; however, the results are similar for GMOS-North data cubes.

\begin{figure}
\begin{center}
  \includegraphics[scale=0.55]{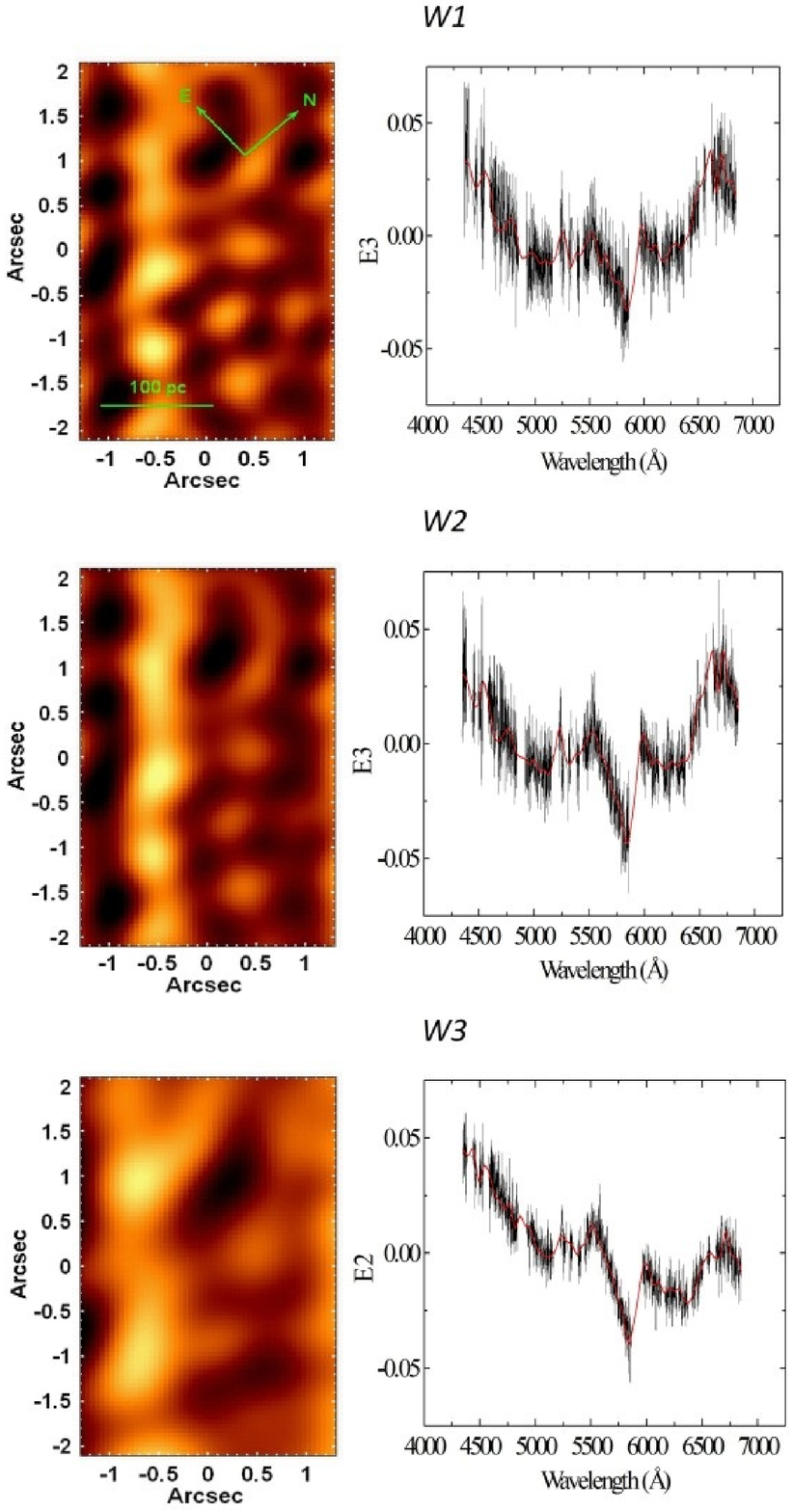}
  \caption{Examples of the eigenspectra, and of the corresponding tomograms, related to the instrumental fingerprint in the data cube of the central region of NGC 1399, provided by the PCA Tomography technique applied to the $W_1$, $W_2$ and $W_3$ data cubes, resulting from the wavelet decomposition. The splines fitted to the eigenspectra are shown in red. The spectral signature and the spatial morphology (with different spatial frequencies) of the instrumental fingerprint can be easily seen.\label{fig14}}
\end{center}
\end{figure}

\begin{figure*}
\begin{center}
  \includegraphics[scale=2.1]{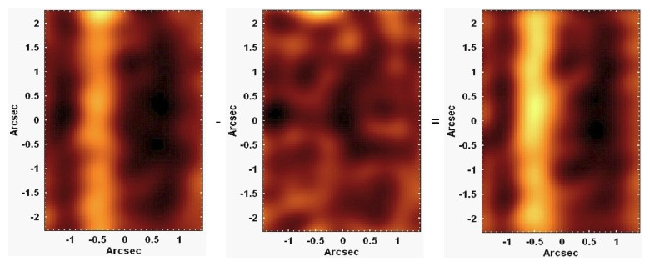}
  \caption{Left: image obtained, from the data cube of the central region of the galaxy NGC 1399 before the instrumental fingerprint removal, by subtracting the image of the wavelength interval $5750 - 5862$ \AA from the image of the wavelength interval $5978 - 6090$ \AA. Centre: the same as in the image on the left, but from the data cube of NGC 1399 after the instrumental fingerprint removal. Right: difference between the two previous images.\label{fig15}}
\end{center}
\end{figure*}

\subsection{Removal with W-PCA Tomography}

In certain situations, we verified that the PCA Tomography technique alone is not able to completely isolate the fingerprint in specific eigenvectors. As a consequence, traces of the fingerprint remain mixed with other physical phenomena in different eigenvectors and end up not being removed. In these situations, it is advisable to use the W-PCA Tomography approach, which combines PCA Tomography with the wavelet decomposition. The first step of this procedure is, again, the removal of the main absorption and emission lines. Then we apply the wavelet decomposition to the data cube without the main spectral lines. After that, the PCA Tomography technique is applied to each wavelet data cube and the eigenvectors associated with the fingerprint are identified. Splines are fitted to the eigenspectra related to the fingerprint and, finally, a data cube containing only the instrumental fingerprint is constructed, using the fitted splines and the corresponding tomograms, and subtracted from the original data cube. Therefore, the difference between the approaches involving PCA Tomography and W-PCA Tomography is that, in the former, PCA Tomography is applied to the data cube without the main spectral lines, while, in the latter, PCA Tomography is applied to the wavelet data cubes resulting from the wavelet decomposition of the data cube without the main spectral lines. The main advantage of W-PCA Tomography is that, by separating spatial structures corresponding to different spatial frequencies, the wavelet decomposition may be able to separate the fingerprint from other physical phenomena. Then, when PCA Tomography is applied to the wavelet data cubes, the fingerprint is well isolated in one or more eigenvectors, which, as mentioned before, may not be possible, in some cases, when PCA Tomography is applied directly to the data cube without the main spectral lines.

We applied the W-PCA Tomography process to remove the instrumental fingerprint of the data cube of the central region of NGC 1399, obtained with GMOS-South, as the PCA Tomography procedure was not able to completely remove the fingerprint from this data cube. Fig.~\ref{fig14} shows some eigenspectra related to the instrumental fingerprint, together with the fitted splines and the associated tomograms, provided by PCA Tomography of the wavelet data cubes of NGC 1399. We can see that PCA Tomography isolated the fingerprint (which was decomposed by the wavelet algorithm in components with different spatial frequencies) in a number of eigenvectors, without significant contaminations by other phenomena. Similarly to the cases of NGC 1316 and NGC 157, the great efficacy of the W-PCA Tomography technique can be seen in the subtraction of the image of the wavelength interval $5750 - 5862$ \AA~from the image of $5978 - 6090$ \AA~(Fig.~\ref{fig15}), performed before and after the instrumental fingerprint removal of the data cube of NGC 1399. W-PCA Tomography produces similar results for GMOS-South and GMOS-North data cubes obtained before and after the CCD replacement.

\begin{figure*}
\begin{center}
  \includegraphics[scale=1.5]{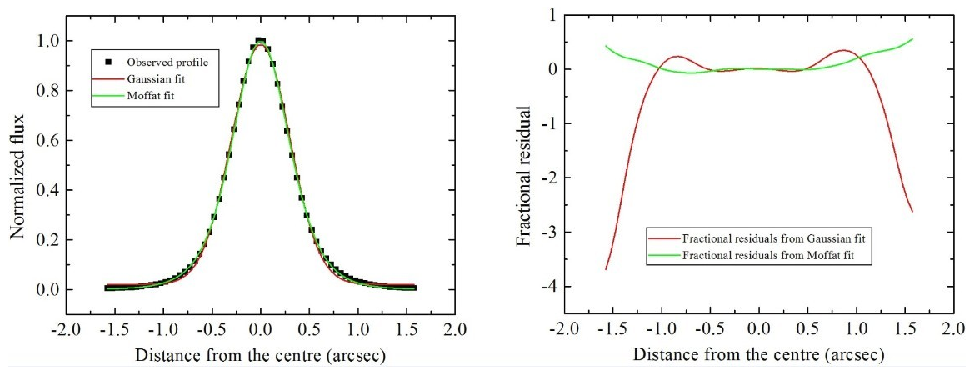}
  \caption{Left: Horizontal brightness profile of an intermediate-wavelength image of the data cube of the star HZ 44. Right: Fractional residuals of the fits shown on the left.\label{fig16}}
\end{center}
\end{figure*}

The cause of the GMOS/IFU instrumental fingerprint is not known yet. Scattered light in the instrument is certainly a possibility that must be taken into account. However, the observed spectral pattern, especially after the CCD replacement, seems to suggest a different origin for this feature. Each GMOS CCD is read out through a number of amplifiers. Bias and gain level differences between such amplifiers would result in this pattern. Although such differences should have been corrected during the data reduction, residual effects not removed by the data reduction seem to be a plausible explanation for the observed instrumental fingerprint in GMOS/IFU data cubes.

\section{Richardson-Lucy deconvolution}

\begin{figure}
\begin{center}
  \includegraphics[scale=2.0]{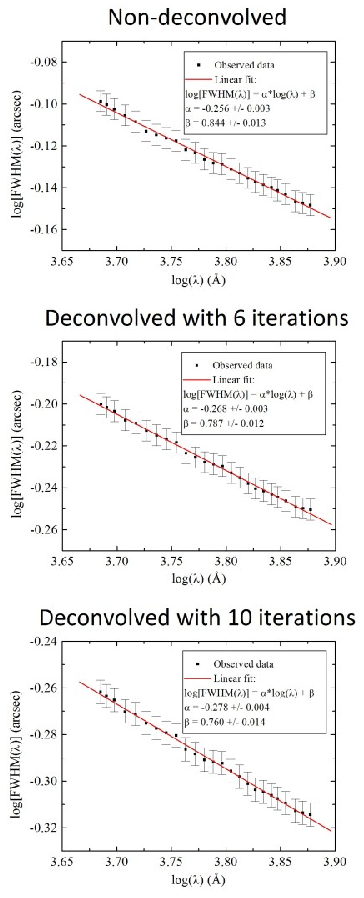}
  \caption{Graphs of the values of $log(FWHM)$ of the PSF as a function of $log(\lambda)$ (with $\lambda$ corresponding to the wavelength, in Angstroms), obtained from the data cube of the star HZ 44 before and after the Richardson-Lucy deconvolution, with 6 and 10 iterations. In each graph, a linear fit is shown in red. The deconvolution reduces the values of FWHM without changing significantly the relation between $FWHM$ and $\lambda$ or the values of $\alpha$.\label{fig17}}
\end{center}
\end{figure}

A deconvolution is an iterative procedure used to obtain the opposite effect of a convolution and can be used in astronomy to improve the spatial resolution of ground-based observations. For more details, see Papers I and II. As the last step in our treatment procedure, we apply the Richardson-Lucy deconvolution \citep{ric72,luc74} to each image of GMOS/IFU data cubes. Although this is the same method we use for the treatment of NIFS and SINFONI data cubes (see Papers I and II), the specific details of the process show significant differences.

The first point that must be considered is related to the profile of the PSF. Fig.~\ref{fig16} shows the horizontal brightness profile of an intermediate-wavelength image of the data cube of HZ 44, together with a Gaussian fit and a Moffat fit. Although these two functions reproduce the main aspects of the profile, the wings of the PSF are better reproduced by a Moffat fit. This can be easily observed in the fractional residuals [(Obs-Fit)/Obs], also shown in Fig.~\ref{fig16}. Therefore, we conclude that, in the case of GMOS/IFU data cubes, the best estimate for the profile of the PSF of the observation, which is a required parameter to perform the Richardson-Lucy deconvolution, is given by a Moffat function. However, our tests revealed that deconvolutions performed assuming Gaussian PSFs also result in good improvements of the spatial resolution of the observations.

The second point that must be evaluated is the way the PSF varies with the wavelength. In Papers I and II, we applied the Richardson-Lucy deconvolution to NIFS and SINFONI data cubes assuming a constant PSF along the spectral axis of the data cube. However, our tests revealed that such a variation of the PSF with the wavelength is more significant in the case of GMOS/IFU data cubes and, therefore, cannot be ignored. Fig.~\ref{fig17} shows a graph with the values of $log$(FWHM) of the PSF as a function of $log(\lambda)$ (with $\lambda$ corresponding to the wavelength, in Angstroms), obtained from the data cube of HZ 44. The values of FWHM were determined by making images of a number of wavelength intervals, with 15 \AA~each, of the data cube of HZ 44 and then by fitting Moffat functions to these images. The graph in Fig.~\ref{fig17} also shows a linear fit, which reproduces very well the behaviour of the observed points. Therefore, we conclude that the values of FWHM($\lambda$) of the PSF of GMOS/IFU data cubes are given by the following relation:

\begin{equation}
FWHM \left(\lambda\right) = FWHM_{ref} \cdot \left(\frac{\lambda}{\lambda_{ref}}\right)^{\alpha},
\end{equation}

where FWHM$_{ref}$ is the value of FWHM at $\lambda_{ref}$. Our tests showed that $\alpha$ is normally in the range of -0.35 to -0.25 for GMOS/IFU data cubes. The average value of $\alpha$ = -0.3 produces good results for most of the situations. Similarly to the cases of NIFS and SINFONI data cubes, we verified that the Richardson-Lucy deconvolution provides the best results with a number of iterations between 6 and 10. A deconvolution performed with less than six iterations will not result in a considerable improvement of the spatial resolution. On the other hand, a number of iterations higher than 10 will cause an amplification of the low-frequency spatial noise. We applied the Richardson-Lucy deconvolution, with 6 and 10 iterations, to the data cube of HZ 44, using a Moffat wavelength-dependent PSF given by equation (1) with $\alpha = -0.256$, and also FWHM$_{ref} = 0.74$ arcsec and $\lambda_{ref} = 6260$ \AA, as estimated from the data cube. The graphs of $log$(FWHM) $\times$ $log(\lambda)$ for the deconvolved data cubes, with 6 and 10 iterations, are shown in Fig.~\ref{fig17}. The deconvolution resulted in FWHM$_{ref} = 0.59$ arcsec and FWHM$_{ref} = 0.51$ arcsec, at $\lambda_{ref} = 6260$ \AA, using 6 and 10 iterations, respectively. Our previous experiences show that the FWHM of the PSF of a deconvolved GMOS/IFU data cube in a given wavelength is $\sim 68\% - 81\%$ of the original value, for a number of iterations between 6 and 10. It is worth mentioning that, if the deconvolution is applied to a GMOS/IFU data cube with larger spaxels (with 0.1 or 0.2 arcsec), the improvements in the spatial resolution are not so high and the FWHM of the final PSF is usually $\sim 85\%$ of the original value. We can also see, in the graphs in Fig.~\ref{fig17}, that the deconvolution does not change significantly the relation between FWHM and $\lambda$ or the values of $\alpha$.

\section{Comparison with HST images}

The improvements in the images of GMOS/IFU data cubes provided by the treatment procedures described in the previous sections are evident. However, at this point, one could ask: Do these treatment processes introduce distortions or even ghost features in the images of the data cubes? Are the spatial structures of the observed object really preserved after the data treatment? In order to answer these questions, we compared images from the GMOS/IFU data cubes of the central regions of the galaxies NGC 1313, NGC 5236 and NGC 7424 with HST images of these same objects. 

\begin{figure*}
\begin{center}
  \includegraphics[scale=2.45]{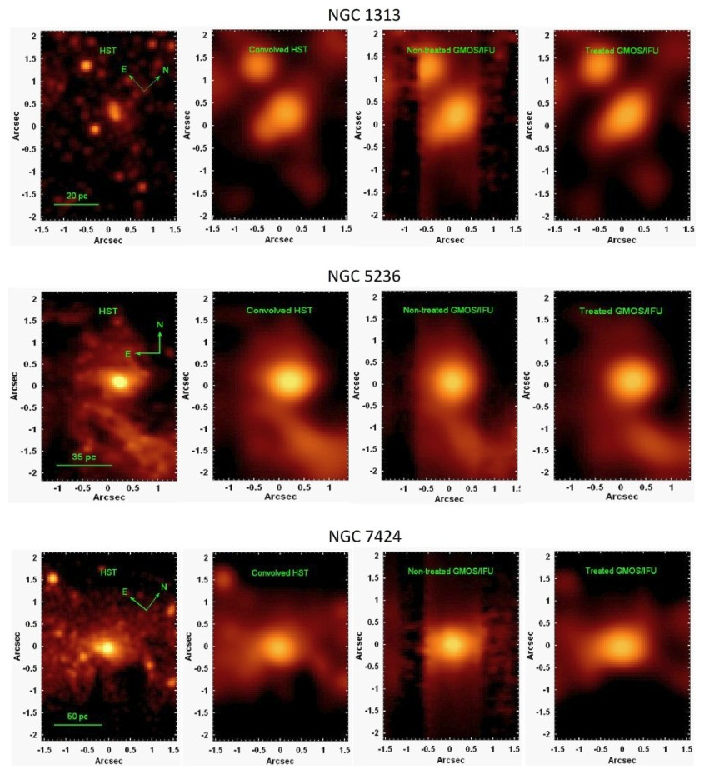}
  \caption{HST images of the central regions of the galaxies NGC 1313, NGC 5236 and NGC 7424 before and after the convolution with the representative PSFs of the corresponding treated GMOS/IFU data cubes. The FOV and orientation of the HST images were taken as the same of the corresponding GMOS/IFU data cubes. In order to establish a comparison, the images of the treated and non-treated GMOS/IFU data cubes of these three galaxies, integrated along the spectral axis, taking into account the transmission curves of the HST filters used for the observations, are also shown.\label{fig18}}
\end{center}
\end{figure*}

All the HST images were retrieved from the HST data archive. NGC 1313 was observed on 2004 July 7, using the Advanced Camera for Surveys (ACS) with the F555W filter. One 680 s exposure was retrieved. NGC 5236 was observed on 2016 January 21, using the Wide Field Camera 3 (WFC3), also with the F555W filter. One 1452 s exposure was retrieved. Finally, NGC 7424 was observed on 1994 July 16, using the Wide Field Planetary Camera 2 (WFPC2) with the F606W filter. In this case, two 80 s exposures were retrieved and an average of the images was calculated.

\begin{figure}
\begin{center}
  \includegraphics[scale=1.4]{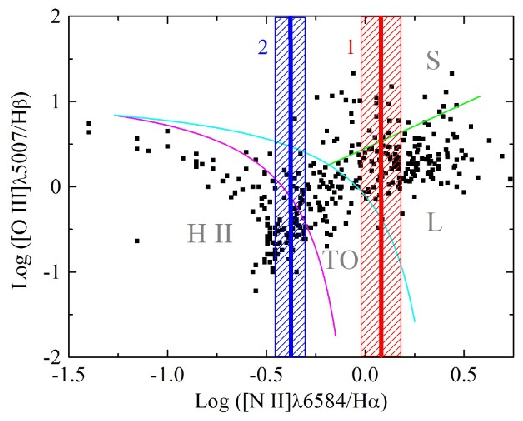}
  \caption{Diagnostic diagram of [O III] $\lambda$5007/H$\beta$ $\times$ [N II] $\lambda$6584/H$\alpha$. The points correspond to the objects analysed by \citet{ho97}. The magenta curve corresponds to an empirical division between H II regions and AGNs determined by \citet{kau03}, the cyan curve represents the maximum limit for the ionization by a starburst obtained by \citet{kew01} and the green curve corresponds to the division between LINERs and Seyferts determined by \citet{sch07}. The regions corresponding to Seyfert galaxies, LINERs, Transition objects and H II regions are represented by S, L, TO, and H II, respectively. The red and blue vertical lines represent the [N II] $\lambda$6584/H$\alpha$ values of the spectra extracted from regions 1 and 2, respectively, in the treated data cube of NGC 2835. The striped area around these vertical lines indicates the uncertainties of the [N II] $\lambda$6584/H$\alpha$ values.\label{fig19}}
\end{center}
\end{figure}

First of all, we integrated the treated and non-treated data cubes of the three objects, along their spectral axes, taking into account the transmission curves of the HST filters used for the observation of each galaxy. For this analysis, we took, as the non-treated data cube, the result obtained after the correction of the DAR and the combination of data cubes in the form of a median. We rotated the HST images to match their orientations with the ones of the data cubes of the corresponding objects. After that, we trimmed the HST images, so that their final FOVs were the same of the corresponding GMOS/IFU data cubes. In order to obtain a representative GMOS/IFU PSF, we constructed, for each galaxy, a data cube containing, in each frame, the Moffat profile corresponding to the PSF in that specific wavelength, before the Richardson-Lucy deconvolution. The PSF profile for each wavelength was obtained taking into account the value of $FWHM_{ref}$ provided by the acquisiton image, taken at $\lambda_{ref} = 6300$ \AA, and also the relation between the FWHM of the PSF and the wavelength (equation 1) provided by the data cube of the standard star used in the data reduction. We then applied the Richardson-Lucy deconvolution to the obtained PSF data cube, with the same parameters used in the deconvolution of the science data cube. Finally, following the same approach mentioned above, we integrated the deconvolved PSF data cube, along its spectral axis, taking into account the transmission curve of the HST filter used for the observation of the corresponding galaxy. The resulting image was taken as the representative PSF of the treated GMOS/IFU data cube of this galaxy. Each representative PSF was then convolved with the rotated and resized HST image of the galaxy. Fig.~\ref{fig18} shows the (rotated and resized) HST images of the galaxies NGC 1313, NGC 5236, and NGC 7424, before and after the convolution with the representative PSFs of the corresponding treated data cubes. The integrated images of the treated and non-treated data cubes of the galaxies are also shown in Fig.~\ref{fig18}. It is easy to see the great consistency between the integrated images of the treated data cubes and the convolved HST images. Structures like apparent spiral arms in NGC 5236 and stellar clusters or even individual stars in NGC 1313 and NGC 7424, which are easily identified in the convolved HST images, also appear clearly in the images of the integrated treated data cubes. In addition, a comparison between the integrated images of the treated and non-treated data cubes reveals the great improvement provided by our treatment procedure. In particular, the effects of the instrumental fingerprint, which compromise considerably the visualization of the spatial structures in the non-treated data cubes, are essentially absent in the integrated images of the treated data cubes. These results certainly indicate that the treatment procedure presented in this work not only improves the quality of the images of the data cubes but also does not compromise the morphology of the spatial structures or introduce possible ghost features.

\begin{figure}
\begin{center}
  \includegraphics[scale=1.5]{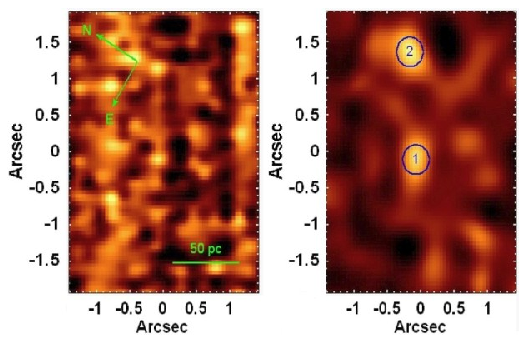}
  \caption{Left: images of the wavelength interval $6581 - 6587$ \AA (corresponding to the [N II] $\lambda$6584 emission line), after the subtraction of an image of the underlying stellar continuum, from the non-treated data cube of the central region of NGC 2835. Right: the same image shown on the left, but from the treated data cube of NGC 2835. The non-treated image shows essentially noise. However, the treated image shows two compact line-emitting regions, marked on the image, from which the spectra shown in Fig.~\ref{fig21} were extracted.\label{fig20}}
\end{center}
\end{figure}

\section{A scientific example: NGC 2835}

The study of the nuclear regions of galaxies is a very important topic, as it can reveal relevant information about the formation and evolution of these objects. While some galaxies have a nuclear spectrum showing only a stellar continuum and absorption lines, other objects show emission lines superimposed to the stellar nuclear spectrum. Besides an appropriate starlight subtraction, the analysis of the nuclear emission-line properties of galaxies usually involves diagnostic diagrams \citep{bal81}, which are essentially graphs of different emission-line ratios that can separate the line-emitting objects in H II regions, Seyfert galaxies and Low Ionization Nuclear Emission-Line Regions (LINERs). An example of a diagnostic diagram of [O III] $\lambda$5007/H$\beta$ $\times$ [N II] $\lambda$6584/H$\alpha$ is shown in Fig.~\ref{fig19}. Seyfert galaxies form a class of the so-called active galactic nuclei (AGNs), which are characterized by emission-line spectra that cannot be attributed to stellar processes. According to their spectral and photometric properties, AGNs are divided into different classes, like quasars, Quasi Stellar Objects (QSOs), Seyfert galaxies, radio galaxies, etc. Today it is well accepted that the energy emitted by AGNs comes from matter accretion onto a central supermassive black hole (SMBH). This mechanism, with a lower ionization parameter, has also been proposed to explain the observed spectral properties of LINERs \citep{fer83,hal83}. Therefore, LINERs are usually interpreted as another class of AGNs. However, it is known that other processes like shock heating \citep{hec80,dop95,dop96}, photoionization by young stars \citep{ter95,fil92,shi92} and photoionization by post-AGB stars \citep{bin94,sta08,cid11,era10} can also generate emission-line ratios typical of LINERs. This suggests that LINERs may be a heterogeneous class of objects.

\begin{figure*}
\begin{center}
  \includegraphics[scale=1.8]{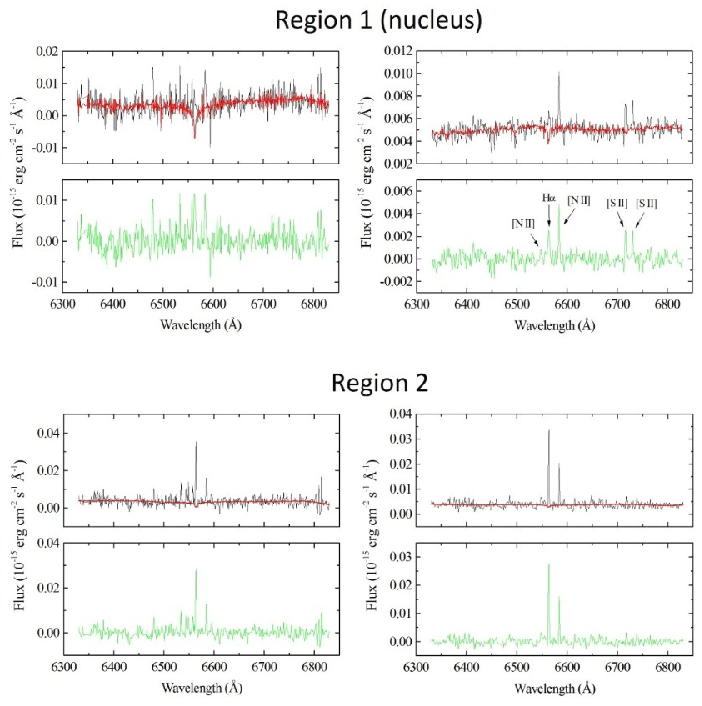}
  \caption{Red portion of the spectra of Regions 1 and 2 extracted from the (left) non-treated and (right) treated data cubes of NGC 2835. The fits provided by the pPXF technique are shown in red and the fit residuals are shown in green. The spectra from the non-treated data cube are noisier and, in the case of Region 1, it is not possible to confirm the detection of any emission line.\label{fig21}}
\end{center}
\end{figure*}

We are conducting the Deep IFS View of Nuclei of Galaxies ($DIVING^{3D}$) survey (PI: J. E. Steiner), whose purpose is to observe, using the GMOS/IFU, the central region of all southern galaxies brighter than B = 12.0. The methodologies described in the previous sections are being used in the treatment of all data cubes obtained for this survey. One of the main goals of the $DIVING^{3D}$ project is to analyse the properties of the nuclear emission-line spectra of all galaxies in the sample. More specifically, we want to determine, for example, the fraction of galaxies, with different morphological types, with a nuclear emission consistent with that of LINERs or Seyferts. \citet{ho08} analysed long-slit spectra of the central regions of all northern galaxies brighter than B = 12.5 (PALOMAR survey). They verified that $\sim $50 - 70 \% of the galaxies with morphological types E - Sb can be classified as Seyferts, LINERs, or Transition objects (which correspond to the objects that fall, on a diagnostic diagram, between the empirical division between H II regions and AGNs of Kauffmann et al. 2003 and the maximum limit for the ionization by a starburst of Kewley et al. 2001 - see Fig.~\ref{fig19}). On the other hand, only $\sim$20 \% of the galaxies with morphological types Sc - Sm show this AGN-type emission. The difficulty of detecting low-luminosity AGNs, however, may affect these fractions. This is especially problematic in the case of late-type galaxies, whose strong emission from H II regions may obfuscate the emission from possible low-luminosity AGNs. Data obtained with IFUs have been used in recent studies to search for hidden nuclear sources (e.g. Belfiore et al. 2016; Wylezalek et al. 2018). By using high spatial and spectral resolution GMOS/IFU data cubes, we believe we will be able to detect considerably faint AGNs and possibly alter the values of the fractions mentioned before. Examples of detections of faint AGNs in GMOS/IFU data cubes, treated with our treatment procedure, can be found in \citet{ric11,men14b,ric14a,ric14b}.

As shown in the previous sections, artefacts like high spatial-frequency noise or instrumental fingerprints may have a significant impact on GMOS/IFU data cubes and, as a consequence, may compromise different analyses performed with these data. One example of such an analysis involves the search for low-luminosity AGNs. In order to show how this type of analysis can be considerably affected by instrumental artefacts, we used the data cube, from the $DIVING^{3D}$ sample, of the central region of the SB(rs)c galaxy NGC 2835, located at a distance of 10.36 Mpc (NASA Extragalactic Database - NED). There is not much information in the literature about this object. The data cube of this galaxy was treated with all the methodologies described before. We corrected the treated and non-treated data cubes of NGC 2835 for the interstellar extinction due to the Galaxy, using $A_V = 0.274$ mag \citep{sch11} and the extinction law of \citet{car89}, and passed them to the rest frame, using $z = 0.002955$ (NED). We also sampled the spectra with $\Delta\lambda = 1$ \AA. Fig.~\ref{fig20} shows images of the wavelength interval $6581 - 6587$ \AA (corresponding to the [N II] $\lambda$6584 emission line), after the subtraction of an image of the underlying stellar continuum, from the treated and non-treated data cubes of NGC 2835. The non-treated image is very noisy and does not show clear line-emitting regions. Traces of large vertical stripes, possibly related to the instrumental fingerprint, can also be seen in the non-treated image. On the other hand, two compact line-emitting regions (Regions 1 and 2, from now on) can be seen in the treated image, one of them (Region 1) coinciding with the stellar nucleus. 

In order to analyse the emission-line spectra of the two emitting regions detected in the treated data cube of NGC 2835, we extracted spectra from circular regions, with a radius of half of the FWHM of the PSF at the median wavelength of the data cube, centred on each one of these regions. Such spectra were extracted from the treated and non-treated data cubes. In the case of the non-treated data cube, the FWHM of the PSF at the median wavelength of the data cube ($FWHM(5832 \AA)_{non-treated} = 0.52$ arcsec) was estimated taking into account the relation between the FWHM of the PSF and the wavelength (Equation 1), with $\alpha = -0.288$, as determined from the data cube of the standard star used for the flux calibration during the data reduction, and FWHM$_{ref} = 0.51$ arcsec, as determined from the acquisition image, with $\lambda_{ref} = 6300$ \AA. On the other hand, in the case of the treated data cube, the FWHM of the PSF at the median wavelength ($FWHM(5832 \AA)_{treated} = 0.35$ arcsec) was taken as $68\%$ of the value determined for the non-treated data cube. This last calculation was performed based on previous results we obtained for the Richardson-Lucy deconvolution of GMOS/IFU data cubes (see Section 6). We then performed a stellar continuum subtraction of the extracted spectra with the Penalized Pixel Fitting procedure (pPXF - Cappellari 2017). This method combines template stellar population spectra from a base, convolved with a Gauss-Hermite expansion, to fit the stellar spectrum of a given object. In this case, we used a base of stellar population spectra based on Medium-resolution Isaac Newton Telescope Library of Empirical Spectra (MILES; S\'anchez-Bl\'azquez et al. 2006). Besides providing stellar kinematic parameters like the stellar radial velocity, the stellar velocity dispersion and the Gauss-Hermite coefficients $h_3$ and $h_4$, this method also results in a synthetic stellar spectrum, corresponding to the fit of the stellar component of the observed spectrum. We subtracted these synthetic stellar spectra provided by pPXF from the spectra extracted from the treated and non-treated data cubes of NGC 2835. 

Fig.~\ref{fig21} shows the red portion of all the extracted spectra, together with the fits provided by pPXF and the fit residuals. The blue portion of these spectra is not shown here because it does not reveal any relevant emitting feature. The fit residuals of the spectrum of Region 1 extracted from the non-treated data cube are very noisy and it is not possible to confirm the detection of any emission line. On the other hand, the fit residuals of the spectrum of this same region extracted from the treated data cube are considerably different and show the H$\alpha$, [N II] $\lambda \lambda$6548,6584 and [S II] $\lambda \lambda$ 6716,6731 emission lines. We obtained an emission-line ratio of [N II] $\lambda$6584/H$\alpha$(1)$_{treated} = 1.2 \pm 0.3$ for this spectrum. As can be seen in Fig.~\ref{fig19}, such ratio indicates the presence of a Transition object or of a LINER or a Seyfert. Unfortunately, without the [O III] $\lambda$5007/H$\beta$ ratio, it is not possible to precisely determine the classification of this object. Even so, based on the low number of Transition Objects detected by \citet{ho97} with [N II] $\lambda$6584/H$\alpha$ values compatible with [N II] $\lambda$6584/H$\alpha$(1)$_{treated}$ and also on the fact that Seyfert nuclei are usually more luminous than the nucleus of NGC 2835, we conclude that this galaxy probably hosts a LINER, although the classifications of Transition Object or Seyfert cannot be ruled out. The H$\alpha$ and [N II] $\lambda \lambda$6548,6584 emission lines are also visible in the fit residuals of the spectra of Region 2 extracted from the treated and non-treated data cubes of NGC 2835. However, the noise in the spectrum from the treated data cube is much lower. The calculated emission-line ratios for the spectra of Region 2 from the treated and non-treated data cubes are [N II] $\lambda$6584/H$\alpha$(2)$_{treated} = 0.42 \pm 0.07$ and [N II] $\lambda$6584/H$\alpha$(2)$_{non-treated} = 0.40 \pm 0.17$, respectively. The two values are compatible, at a 1$\sigma$ level, but, as expected, the reduced noise in the spectrum from the treated data cube resulted in a lower uncertainty for the emission-line ratio. As Fig.~\ref{fig19} shows, this emission-line ratio is consistent with that of an H II region, a Transition object or even a Seyfert galaxy. Based on the low luminosity of this emitting region and also on the low number of Seyfert galaxies observed by \citet{ho97} with [N II] $\lambda$6584/H$\alpha$ values compatible with [N II] $\lambda$6584/H$\alpha$(2)$_{treated}$, we conclude that the most likely classifications for Region 2 are H II region or Transition Object.

The previous results show that, without our data treatment, it would not have been possible to detect a Seyfert/LINER/Transition object emission in the nucleus of NGC 2835. The same would probably happen with other objects with faint emission lines. This problem would affect any analysis of GMOS/IFU data cubes focused on the determination of the fraction of galaxies, with different morphological types, with a central Seyfert/LINER/Transition object emission. Considering that, we conclude that the treatment procedure we presented in this work not only improves the appearance of the images of GMOS/IFU data cubes, but actually removes a number of artefacts that could potentially compromise different types of analyses.

\section{Summary and conclusions}

We presented a set of treatment techniques to be applied to GMOS/IFU data cubes, including correction of the DAR, Butterworth spatial filtering, instrumental fingerprint removal, and Richardson-Lucy deconvolution. This treatment procedure has the purpose of improving the quality of the data by removing artefacts, like high spatial-frequency noise and the instrumental fingerprint, and also by improving the spatial resolution of the observation. The main conclusions of this work are:

\begin{itemize}

\item The spatial displacements, along the spectral axis, caused by the DAR effect in GMOS/IFU data cubes can be comparable or even larger than the typical spatial resolutions of this instrument. Therefore, this effect can compromise the analysis of the data and must be removed. The proposed practical and theoretical approaches for removing the DAR are both very efficient and result in precisions between 1 and 10 mas for the final positions of the spatial structures along the spectral axis of the data cubes.

\item GMOS/IFU data cubes show a high spatial-frequency noise whose morphology suggests that it is related to the lenslet/fibre working mechanism of the instrument. Part of the noise is also introduced by the spatial resampling, when the data cube is reduced with spaxels smaller than 0.2 arcsec. A Butterworth spatial filtering, with a squared circular filter with order $n = 2$ and cut-off frequencies between 0.16 and 0.25 Ny, removes most of the high spatial-frequency noise, without introducing artefacts or degrading the PSF.

\item Similarly to NIFS and SINFONI data cubes, GMOS/IFU data cubes often show an instrumental fingerprint, with a low-frequency spectral signature, that takes the form of vertical stripes across the images. In most cases, our removal technique, with PCA Tomography, can efficiently isolate and remove the fingerprint. However, in certain situations, when PCA Tomography alone is not able to completely isolate the fingerprint from other physical phenomena in the data cube, it is advisable to remove the fingerprint with the W-PCA Tomography approach, which combines PCA Tomography and the wavelet decomposition. 

\item The profile of the PSF of GMOS/IFU data cubes takes the form of a Moffat function, although a Gaussian function also reproduces the main aspects of such a profile. The FWHM of the PSF varies with the wavelength according to Equation (1), with $\alpha$ normally in the range of -0.35 to -0.25. The Richardson-Lucy deconvolution, with a number of iterations between 6 and 10 and assuming a Moffat or even a Gaussian function, improves considerably the spatial resolution of a GMOS/IFU data cube, resulting in a PSF with an FWHM $\sim 68\% - 81\%$ of the original value. The deconvolution does not change significantly the relation between the FWHM of the PSF and the wavelength.

\item A comparison with HST images shows that the treatment procedure does not introduce distortions or ghost features in the images of GMOS/IFU data cubes.

\item The use of the treatment procedure in the data cube of the central region of the late-type galaxy NGC 2835 allowed the detection of a nuclear emitting region, whose emission-line spectrum is probably of a LINER, although the classifications of Transition Object or Seyfert galaxy cannot be discarded. Without the data treatment, such a detection was not possible, due to the fact that the faint emission lines were immersed in a considerable spectral noise. This result shows that our treatment procedure not only improves the appearance of the images of GMOS/IFU data cubes, but actually removes a number of artefacts that could potentially compromise different types of analyses performed with the data cubes.

\end{itemize}

The scripts of all treatment techniques described in this work can be found at \textit{http://www.astro.iag.usp.br/$\sim$PCAtomography}.

\section*{Acknowledgements}

Based on observations obtained at the Gemini Observatory (processed using the Gemini IRAF package), which is operated by the Association of Universities for Research in Astronomy, Inc., under a cooperative agreement with the NSF on behalf of the Gemini partnership: the National Science Foundation (United States), the National Research Council (Canada), CONICYT (Chile), the Australian Research Council (Australia), Minist\'{e}rio da Ci\^{e}ncia, Tecnologia e Inova\c{c}\~{a}o (Brazil) and Ministerio de Ciencia, Tecnolog\'{i}a e Innovaci\'{o}n Productiva (Argentina). This work was supported by CAPES (PNPD - PPG-F\'isica - UFABC), CNPQ (under grant 304321/2016), and FAPESP (under grant 2011/51680-6).


\begin{thebibliography}{}

\bibitem[Allington-Smith et al.(2002)]{all02} Allington-Smith J., et al., 2002, PASP, 114, 892
\bibitem[Bacon et al.(2001)]{bac01} Bacon R., et al., 2001, MNRAS, 326, 23
\bibitem[Baldwin, Phillips \& Terlevich(1981)]{bal81} Baldwin J.A., Phillips M.M., Terlevich R., 1981, PASP, 93, 5
\bibitem[Belfiore et al.(2016)]{bel16} Belfiore F., et al., 2016, MNRAS, 461, 3111
\bibitem[Binette et al.(1994)]{bin94} Binette L., Magris C.G., Stasinska G., Bruzual A.G., 1994, A\&A, 292, 13
\bibitem[Bonnet et al.(2004)]{bon04} Bonnet H., et al., 2004, The Messenger, 117, 17
\bibitem[B\"onsch \& Potulski(1998)]{bon98} B\"onsch G., Potulski E., 1998, Metrologia, 35, 13
\bibitem[Cappellari(2017)]{cap17} Cappellari M., 2017, MNRAS, 466, 798
\bibitem[Cardelli, Clayton \& Mathis(1989)]{car89} Cardelli J.A., Clayton G.C., Mathis J.S., 1989, ApJ, 345, 245
\bibitem[Cid Fernandes et al.(2010)]{cid11} Cid Fernandes R., Stasinska G., Mateus A., Asari N.V., 2011, MNRAS, 413, 1687
\bibitem[Dopita \& Sutherland(1995)]{dop95} Dopita M.A., Sutherland R.S., 1995, ApJ, 455, 468
\bibitem[Dopita \& Sutherland(1996)]{dop96} Dopita M.A., Sutherland R.S., 1996, ApJS, 102, 161
\bibitem[Eisenhauer et al.(2003)]{eis03} Eisenhauer F., et al., 2003, SPIE, in Masanori I., Alan F. M. M., eds, Proc. SPIE Conf. Ser. Vol. 4841, Instrument Design and Performance for Optical/Infrared Ground-based Telescopes. SPIE, Bellingham, p. 1548
\bibitem[Eracleous, Hwang \& Flohic(2010)]{era10} Eracleous M., Hwang J.A., Flohic H.M.L.G., 2010, ApJ, 711, 796
\bibitem[Ferland \& Netzer(1983)]{fer83} Ferland G.J., Netzer H., 1983, ApJ, 264, 105
\bibitem[Filippenko(1982)]{fil82} Filippenko A.V., 1982, PASP, 94, 715
\bibitem[Filippenko \& Terlevich(1992)]{fil92} Filippenko A.V., Terlevich R., 1992, ApJ, 397, L79
\bibitem[Fukunaga(1990)]{fuk90} Fukunaga K., 1990, Statistical Pattern Recognition, 2$^{nd}$ edition, Academic Press, New York
\bibitem[Gonzalez \& Woods(2002)]{gon02} Gonzalez R.C., Woods R.E., 2002, Digital Image Processing, 2$^{nd}$ edition, Prentice-Hall, Englewood Cliffs, NJ
\bibitem[Halpern \& Steiner(1983)]{hal83} Halpern J.P., Steiner J.E., 1983, ApJ, 269, L37
\bibitem[Heckman(1980)]{hec80} Heckman T.M., 1980, A\&A, 87, 152
\bibitem[Ho(2008)]{ho08} Ho L.C., 2008, ARA\&A, 46, 475
\bibitem[Ho, Filippenko \& Sargent(1997)]{ho97} Ho L.C., Filippenko A.V., Sargent W.L.W. 1997, ApJS, 112, 315
\bibitem[Kauffmann et al.(2003)]{kau03} Kauffmann G., et al., 2003, MNRAS, 346, 1055
\bibitem[Kewley et al.(2001)]{kew01} Kewley L.J., Dopita M.A., Sutherland R.S., Heisler C.A., Trevena J., 2001, ApJ, 556, 121
\bibitem[Lucy(1974)]{luc74} Lucy L.B., 1974, AJ, 79, 745
\bibitem[McGregor et al.(2003)]{mcg03} McGregor P., et al., 2003, in Masanori I., Alan F. M. M., eds, Proc. SPIE Conf. Ser. Vol. 4841, Instrument Design and Performance for Optical/Infrared Ground-based Telescopes, SPIE, Bellingham, p. 178
\bibitem[Menezes et al.(2015)]{men15} Menezes R.B., da Silva Patr\'icia, Ricci T.V., Steiner J.E., May D., Borges B.W., 2015, MNRAS, 450, 369 (Paper II)
\bibitem[Menezes, Steiner \& Ricci(2014a)]{men14a} Menezes R.B., Steiner J.E., Ricci T.V., 2014, MNRAS, 438, 2597 (Paper I)
\bibitem[Menezes, Steiner \& Ricci(2014b)]{men14b} Menezes R.B., Steiner J.E., Ricci T.V., 2014b, ApJ, 796, L13
\bibitem[Murtagh \& Heck(1987)]{mur87} Murtagh F., Heck A., 1987, Multivariate Data Analysis, Reidel, Dordrecht
\bibitem[Ricci, Steiner \& Menezes(2011)]{ric11} Ricci T.V., Steiner J.E., Menezes R.B., 2011, ApJ, 734, L10
\bibitem[Ricci, Steiner \& Menezes(2014a)]{ric14a} Ricci T.V., Steiner J.E., Menezes R.B., 2014a, MNRAS, 440, 2419
\bibitem[Ricci, Steiner \& Menezes(2014b)]{ric14b} Ricci T.V., Steiner J.E., Menezes R.B., 2014b, MNRAS, 440, 2442
\bibitem[Richardson(1972)]{ric72} Richardson W.H., 1972, J. Opt. Soc. Am., 62, 55
\bibitem[Riffel et al.(2011)]{rif11} Riffel R., Riffel, Rogemar A., Ferrari F., Storchi-Bergmann T., 2011, MNRAS, 416, 493 
\bibitem[S\'anchez et al.(2012)]{san12} S\'anchez S.F., et al., 2012, A\&A, 538, 8
\bibitem[S\'anchez-Bl\'azquez et al.(2006)]{san06} S\'anchez-Bl\'azquez P., et al. 2006, MNRAS, 371, 703
\bibitem[Schawinski et al.(2007)]{sch07} Schawinski K., et al. 2007, MNRAS, 382, 1415 
\bibitem[Schlafly \& Finkbeiner(2011)]{sch11} Schlafly E.F., Finkbeiner D.P., 2011, ApJ, 737, 103
\bibitem[Shields(1992)]{shi92} Shields J.C., 1992, ApJ, 399, L27
\bibitem[Starck \& Murtagh(2006)]{sta06} Starck J.L., Murtagh F., 2006, Astronomical Image and Data Analysis, 2$^{nd}$ edition, Springer-Verlag, Berlin
\bibitem[Stasinska et al.(2008)]{sta08} Stasinska G., et al. 2008, MNRAS, 391, L29
\bibitem[Steiner et al.(2009)]{ste09} Steiner J.E., Menezes R.B., Ricci T.V., Oliveira A.S., 2009, MNRAS, 395, 64
\bibitem[Terlevich \& Melnick(1985)]{ter85} Terlevich R., Melnick J., 1985, MNRAS, 213, 841
\bibitem[van Dokkum(2001)]{van01} van Dokkum P.G., 2001, PASP, 113, 1420 
\bibitem[Wylezalek et al.(2018)]{wyl18} Wylezalek D., et al., 2018, MNRAS, 474, 1499


\end{thebibliography}
\end{document}